**Extreme scaling of the metadynamics of paths algorithm on the pre-exascale JUWELS Booster supercomputer**

Nitin Malapally[1,2], Marta Devodier[1,3,4], Giulia Rossetti[1,2,5], Paolo Carloni[1,6], Davide Mandelli[1*]

[1]Computational Biomedicine (INM-9), Forschungszentrum Jülich, Wilhelm-Johnen-Straße, 52428, Jülich, Germany
[2]Jülich Supercomputing Centre (JSC), Forschungszentrum Jülich, Wilhelm-Johnen-Straße, 52428, Jülich, Germany
[3]Department of Physics, RWTH Aachen University, 52062 Aachen, Germany
[4]Department of Applied Physics, Science for Life Laboratory, KTH Royal Institute of Technology, SE-171 21 Solna, Sweden
[5]Department of Neurology, University Hospital Aachen (UKA), RWTH Aachen University, Aachen, Germany
[6]Neuroscience and Neuroimaging (INM-11), Forschungszentrum Jülich, Wilhelm-Johnen-Straße, 52428, Jülich, Germany
*corresponding author: d.mandelli@fz-juelich.de

**ABSTRACT**
Molecular dynamics (MD)-based path sampling algorithms are a very important class of methods used to study the energetics and kinetics of rare (bio)molecular events. They sample the highly informative but highly unlikely reactive trajectories connecting different metastable states of complex (bio)molecular systems. The metadynamics of paths (MoP) method proposed by Mandelli, Hirshberg, and Parrinello [Pys. Rev. Lett. 125 2, 026001 (2020)] is based on the Onsager-Machlup path integral formalism. This provides an analytical expression for the probability of sampling stochastic trajectories of given duration. In practice, the method samples reactive paths via metadynamics simulations performed directly in the phase space of all possible trajectories. Its parallel implementation is in principle infinitely scalable, allowing arbitrarily long trajectories to be simulated. Paving the way for future applications to study the thermodynamics and kinetics of protein-ligand (un)binding, a problem of great pharmaceutical interest, we present here the efficient implementation of MoP in the HPC-oriented biomolecular simulation software GROMACS. Our benchmarks on a membrane protein (150,000 atoms) show an unprecedented weak scaling parallel efficiency of over 70% up to 3500 GPUs on the pre-exascale JUWELS Booster machine at the Jülich Supercomputing Center.

## Introduction

Metadynamics of paths (MoP) [1] is an enhanced-sampling method that combines the dynamical algorithm of transition path-sampling, first proposed by Dellago et al. [2], with metadynamics [3], [4], which is an exact [5], [6] free energy method based on the use of collective variables (CVs).
MoP provides a way to sample reactive trajectories connecting metastable states, allowing one to study the important microscopic mechanisms underlying complex molecular transformations, including the calculation of kinetic rates [1]. MoP can also be used to design optimal data-driven

CVs for free energy calculations [7]. So far, however, the method has only been applied to small toy systems using its original implementation (see references [1], [7], [8]) in the LAMMPS software [9], [10].
To pave the way for applications of MoP in the realm of complex biological systems, we have here implemented the algorithm within the HPC-oriented classical biomolecular simulation code GROMACS [11], [12], [13]. In this report, we present the implementation of the MoP algorithm in GROMACS and demonstrate its weak scaling up to 3,500 GPUs on the pre-exascale JUWELS Booster machine[14], using as a benchmark a membrane protein of pharmacological relevance.
The paper is organized as follows. Section 1 reviews the theory of MoP, which is necessary to understand its parallel implementation. Section 2 gives an overview of the workflow of its GROMACS implementation. Section 3 presents the results of our benchmark simulations. Finally, Section 4 draws our conclusions and provides an outlook.

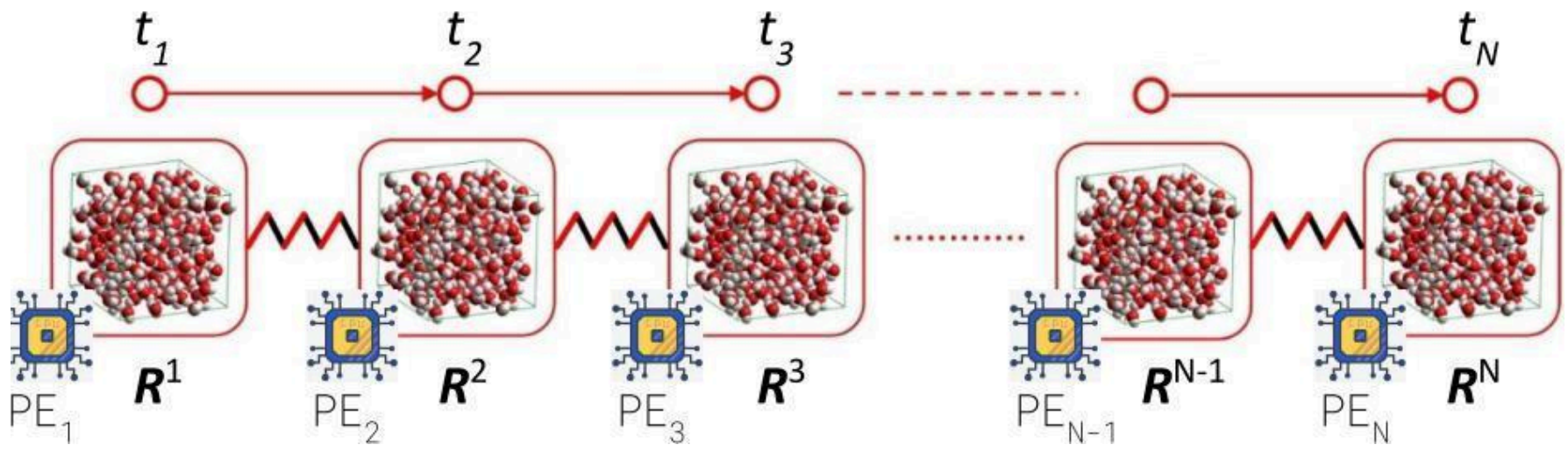

**Figure 1:** The path molecular dynamics algorithm maps a discretized trajectory of *N* steps into an elastic polymer made of *N* beads (*N* replicas of the physical system) that interact via springs. In its parallel implementation, each bead is assigned to an independent processing element (PE) that typically runs on independent hardware resources.

## 1. Theory

### 1.1 Path Molecular Dynamics for Stochastic Trajectories

We consider the dynamics of a molecular system coupled to a thermal bath at temperature $T$. The system contains $Q$ atoms, described by cartesian coordinates $\mathbf{R} = \{\mathbf{r}_1, \ldots, \mathbf{r}_Q\} \in \mathbb{R}^{3Q}$, interacting via a potential $U(\mathbf{R})$. We further assume that the dynamics is governed by the Smoluchowski equation

$$M\nu\dot{\mathbf{R}} = \mathbf{F} + \xi \quad (1)$$

Where $\mathbf{F} = -\nabla_{\mathbf{R}} U$ is the physical force, $\dot{\mathbf{R}}$ is the velocity, $\nu$ is a friction coefficient, $M$ is the diagonal mass tensor and $\xi$ is a white noise term.
Under this assumption, the probability of observing a given discretized trajectory, $\mathbf{R}^1 \to \mathbf{R}^2 \to \cdots \to \mathbf{R}^N$, starting in a metastable state $\mathbf{R}^1$ and ending in $\mathbf{R}^N$ after $N$ time steps $\Delta t$ (e.g., $\Delta t = 1 - 2$ fs), takes the form of a Boltzmann distribution in the enlarged phase space spanned by the N configurations [1], [2]:

$$p(\mathbf{R}^1, \ldots, \mathbf{R}^N) \propto \exp\left[-\beta V_{\text{eff}}(\mathbf{R}^1, \ldots, \mathbf{R}^N)\right] \qquad (2)$$

where $\beta = 1/k_B T$ and the effective potential is given by

$$V_{\text{eff}} = U(\mathbf{R}^1) + \sum_{n=1}^{N-1} \frac{K}{2}(\mathbf{R}^{n+1} - \mathbf{R}^n - \mathbf{L}^n)^2 \quad (3)$$

$V_{\text{eff}}$ is isomorphic to the potential energy of a fictitious polymer, where the beads $\mathbf{R}^n$ are the configurations sequentially visited by the system along the discretized trajectory (see Figure 1). The beads interact via springs of constant $K = \frac{m\nu}{2\Delta t}$ and equilibrium length $\mathbf{L}^n = \frac{m\nu}{\Delta t}\mathbf{F}^n$, where $\mathbf{F}^n$=$\mathbf{F}(\mathbf{R}^n) = -\nabla_{\mathbf{R}} U(\mathbf{R})|_{\mathbf{R}=\mathbf{R}^n}$ is the physical force acting on the *n*th bead.
Equations (2), (3) effectively turn the dynamical problem of finding all possible trajectories $\mathbf{R}^1 \to \mathbf{R}^2 \to \cdots \to \mathbf{R}^N$ that are solutions of equation (1) into a static problem of sampling all the possible polymer configurations $\{\mathbf{R}^1, \ldots, \mathbf{R}^n\}$ distributed according to equation (2). This problem, in turn, can be solved using standard, finite temperature MD approaches, that is, by evolving the dynamics of the polymer according to Newton's equations of motion

$$M\ddot{\mathbf{R}}^n = \mathbf{F}^n_{\text{eff}} \quad (4)$$

coupled to a thermostat, as done for instance in path-integral MD [7]. In equation (4), $M$ and $\ddot{\mathbf{R}}^n$ are the fictitious mass tensor and acceleration of the *n*th bead, $\mathbf{F}^n_{\text{eff}} = -\nabla_{\mathbf{R}^n} V_{\text{eff}}$, is the effective force acting on it, and $n = 1, \ldots, N$. The expression of the force acting on the *n*th bead (1<n<N) is given by

$$\mathbf{F}^n_{\text{eff}} = -K\left(2\mathbf{R}^n - \mathbf{R}^{n-1} - \mathbf{R}^{n+1}\right) + \frac{1}{2}\left(\mathbf{F}^{n-1} - \mathbf{F}^n\right) + \left[\mathbf{F}(\mathbf{R}^n + \varepsilon\eta^n) - \mathbf{F}(\mathbf{R}^n - \varepsilon\eta^n)\right]/4\varepsilon \quad (5)$$

Where the last term is a finite-difference approximation used to avoid expensive calculations of second derivatives of the potential energy $U(\mathbf{R})$ [8], [9], $\eta^n = \mathbf{R}^{n+1} - \mathbf{R}^n - \mathbf{L}^n$, and $\varepsilon$ is a number small enough to guarantee energy conservation in microcanonical simulations. At each step of the dynamics generated by solving numerically equation (4), a new polymer configuration is obtained, which corresponds to another possible discretized trajectory of the system.
For completeness, we provide here also the expressions of the effective force acting on the first (*n=1*) and on the last bead (*n=N*), which represent special cases:

$$\mathbf{F}^1_{\text{eff}} = \frac{1}{2}\mathbf{F}^1 + K\left(\mathbf{R}^2 - \mathbf{R}^1\right) + \left[\mathbf{F}(\mathbf{R}^1 + \varepsilon\eta^1) - \mathbf{F}(\mathbf{R}^1 - \varepsilon\eta^1)\right]/4\varepsilon \quad (6)$$

$$\mathbf{F}^N_{\text{eff}} = -K\left(\mathbf{R}^N - \mathbf{R}^{N-1}\right) + \frac{1}{2}\mathbf{F}^{N-1} \quad (7)$$

From now on, we will refer to equation (4) as path MD (PMD), to distinguish it from the metadynamics of path (MoP) algorithm – described in the next section – where the dynamics of the polymer is modified by the external metadynamics potential.

### 1.2 Metadynamics of Paths

Among all the possible trajectories of given duration, we are interested in those connecting two metastable basins A and B: for example, reactants and products in chemical reactions or bound and unbound states of a molecular complex. The ensemble of all these trajectories form the so-called transition path ensemble that can be analyzed to obtain microscopic information on the molecular mechanism of the reaction as well as to compute kinetic rates [1], [2].
Trajectories belonging to the transition path ensemble are characterized by vanishingly small values of their Boltzmann weight (equation (2)). In other words, sampling such polymer configurations during unbiased PMD simulations is a rare event. To overcome this problem, one can make use of any MD-based enhanced-sampling technique designed to accelerate phase space exploration. Among them, we focus here on metadynamics[3], [4], [5], [6].
Briefly, in a metadynamics simulation, an external potential $V$ is built on-the-fly as a sum of Gaussian. The metadynamics bias $V(s_1, s_2, \dots)$ is constructed as a function of one or more appropriately defined collective variables (CVs) $\{s_i\}$. The latter are functions, $s_i(\mathbf{R})$, of the system's coordinates and should be chosen to best approximate the slow degrees of freedom governing the transformation of interest. The effect of the metadynamics bias is to enhance the fluctuations of such CVs, making it easier for the system to diffuse over the high free-energy barriers separating the targeted metastable states [10]. Most importantly, exact reweighting techniques[11], [12], [13], [14], [15], [16], [17] can be used to recover the correct (unbiased) Boltzmann statistics.
To apply metadynamics in the context of PMD, one needs to define CVs that are able to enhance the sampling of polymer configurations representing trajectories that are part of the transition path ensemble. In our previous publications[1], [18], we have shown that a good CV for this task is provided by a generalized end-to-end polymer distance, expressed as the difference

$$S_{\rm e2e} = s(\mathbf{R}^N) - s(\mathbf{R}^1) \quad (8)$$

between the value of an order parametre $s(\mathbf{R})$ evaluated in the last (*n=N*) and in the first (*n=1*) bead of the polymer. This order parameter must be able to distinguish between the starting and the final states, i.e., $s_{\rm A} \neq s_{\rm B}$, where $s_{\rm A,B}$ are the values taken for $\mathbf{R} \in A, B$. If this condition is met, non-reactive trajectories in which the system visits only one of the two targeted metastable states will correspond to values of $S_{\rm e2e} \approx 0$. On the other hand, reactive trajectories that sample transition events will correspond to values of $|S_{\rm e2e}| \approx |s_{\rm B} - s_{\rm A}|$. By enhancing the fluctuations of the generalized end-to-end distance, the metadynamics bias increases the chances of sampling the reactive trajectories of interest.

## 2. GROMACS implementation of the PMD algorithm

We have implemented the PMD algorithm in the open source GROMACS code[19] (version 2024.0). GROMACS is among the most widely used HPC-oriented biomolecular simulations codes, and amongst the fastest. Its authors aim to provide the highest possible performance and efficiency on any hardware. A native heterogeneous parallelization setup is implemented, using both CPUs and GPUs, with the possibility to offload all force components to GPUs in single precision calculations. The most recent versions also introduced new direct GPU–GPU communication and extended GPU integration, enabling excellent performances across multiple GPUs and efficient multi-node parallelization[20], [21].
The PMD algorithm has been implemented using a multiple program multiple data (MPMD) approach where each bead of the polymer is assigned to a separate processing element (see Figure 1). Within this approach, the various contributions to the fictitious force (equations (5),(6),(7)) are computed by *N* instances of the GROMACS engine that run concurrently on independent resources. Hence, combined with the PMD algorithm, GROMACS is ready to fully exploit modern accelerated massively parallel architectures.

In the following subsections, we briefly discuss the procedure implemented for the computation of the effective forces, the interface to the PLUMED library[22], [23] developed to enable metadynamics simulations, and we present the results of constant energy and constant temperature PMD simulations validating the implementation.

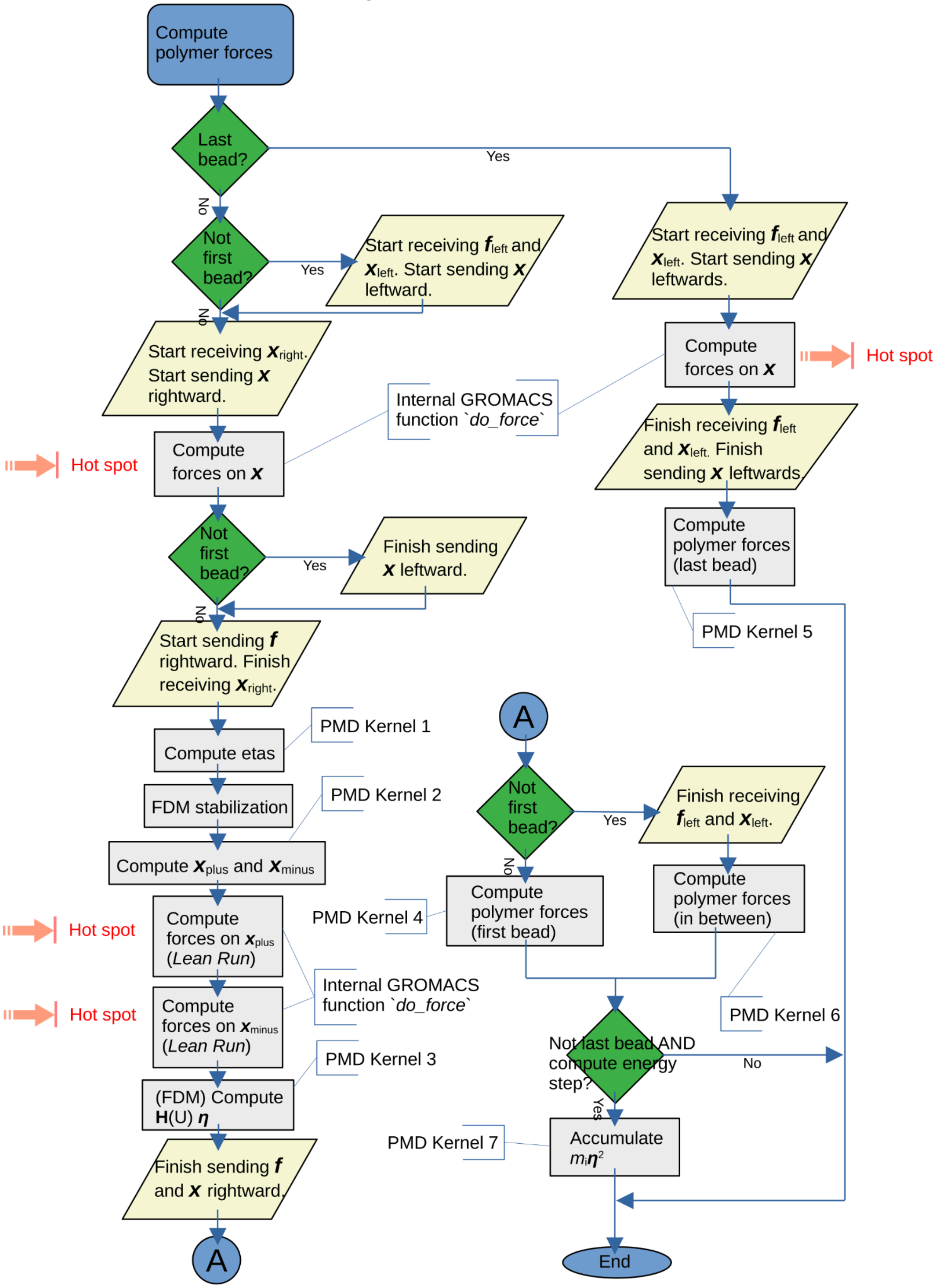

**Figure 2:** Flowchart of our GROMACS implementation of equations (5),(6) and (7). Grey boxes indicate computations, while yellow parallelepipeds indicate inter-replica communication events.

Finite difference method (FDM) stabilization indicates the method by Kapil et al.[9], which we implemented to stabilize the computation of the finite difference expression appearing in equation (5). ***X*** indicates the coordinates $\mathbf{R}^n$ of the current bead, $\boldsymbol{X}_{plus,minus}$ indicate the displaced coordinates, $\boldsymbol{X}_{left,right}$ indicate the coordinates of the adjacent *n-1* and *n+1* bead.

## 2.1 Computing the effective forces

Figure 2 shows a detailed flowchart of the new "*computePolymerForces*" GROMACS' subroutine we have implemented to perform the calculation of the effective forces of equations (5), (6) and (7). Within this function, the majority of the computational effort is concentrated in three distinct calls to GROMACS' internal "*do_force*" function, which computes the physical forces, $\mathbf{F}^n$, acting on a bead. The remaining work is handled by seven newly implemented PMD-specific kernels performing the following operations:

1. Computation of the displacement vectors $\eta^n = \mathbf{R}^{n+1} - \mathbf{R}^n - \mathbf{L}^n$
2. Displacement of the coordinates: $\mathbf{R}^n_{\text{plus,minus}} = \mathbf{R}^n \pm \varepsilon\eta^n$
3. Computation of the finite difference expression $\left[\mathbf{F}^n(\mathbf{R}^n_{\text{plus}}) - \mathbf{F}^n(\mathbf{R}^n_{\text{minus}})\right]/4\varepsilon$
4. Computation of the effective force on the first bead, $\mathbf{F}^1_{\text{eff}}$ (special case)
5. Computation of the effective force on the last bead, $\mathbf{F}^N_{\text{eff}}$ (special case)
6. Computation of the effective force on the nth bead, $\mathbf{F}^n_{\text{eff}}\ 1 < n < N$
7. Computation of the total polymer spring energy (second term in equation (3))

We have optimized these seven kernel functions to make use of OpenMP-based vectorization and multi-threading. For the communication between adjacent beads, required to compute the first two "spring-like" terms in equation (5), we have used non-blocking send and receive operations provided by the MPI library, which avoid additional communication overhead.
After the effective force has been computed, the simulation can proceed as in any standard MD simulation. External forces, for example the metadynamics force, are added, constraints are enforced, *etc.*. The PMD code has been implemented within the legacy code of GROMACS. This allows the user to select any of the available time integration, thermostatting schemes, as well as constraints solver already implemented in GROMACS.
Support for using GROMACS' main parallelization strategies for the computation of the physical forces has also been implemented, including the possibility to use domain decomposition and GPU offloading. These can also be used in combination for multi-GPU offloading. Our implementation can therefore run efficiently both on standard CPU nodes (using domain decomposition) as well as on accelerated nodes exploiting the native GPU offloading features of GROMACS.

## 2.2 PLUMED interface for metadynamics simulations

To enable metadynamics simulations within a PMD simulation, we have modified the existing patch for GROMACS 2024 that is provided by the PLUMED community[1]. Furthermore, we have implemented a dedicated version of PLUMED's CUSTOM function[2] to allow the definition of CVs that depend on the coordinates of different beads, which is necessary to define, *e.g.*, the end-to-end distance of equation (8). In order to minimize the number of inter-replica communication events during a MoP simulation, which can affect both the weak scaling parallel efficiency and overall performance, we have optimized our CUSTOM function in order to ensure that communication occurs only between the replicas involved in the definition of the CV defined by the user.

[1] See https://www.plumed.org/doc-v2.9/user-doc/html/_installation.html
[2] See the PLUMED manual at https://www.plumed.org/doc-v2.9/user-doc/html/_c_u_s_t_o_m.html for a description of the default PLUMED implementation of the CUSTOM function.

### 2.3 NVE and NVT tests

The implementation of the algorithm has been validated by checking energy conservation and stability of the temperature in NVE and NVT simulations, respectively. Several physical systems of increasing complexity have been considered, including a cluster of Lennard-Jones particles, water boxes of different sizes, and a small protein (Lysozyme) in explicit solvent. A book chapter discussing in full detail the results of all the tests, further algorithmic details as well as a thorough performance analysis is in preparation [24]. Here, we focus only on the largest system used so far in our benchmarks, which consists in a model of the human adenosine receptor type 2A in complex with its high affinity antagonist ZMA, embedded in a lipid membrane and solvated in explicit water. The final simulation box contains a total of ~150,000 atoms (see Figure 3). To test the PMD algorithm we used a polymer made of 512 beads, corresponding to a total number of 512×150,000~76.8 Matoms. Further details on the simulations' setup are reported in Appendix A. Figure 4(a) and 4(b) demonstrates, respectively, energy conservation and stability of the system's temperature in short NVE and NVT tests performed on the JUWELS Cluster at the Juelich supercomputing center [25].

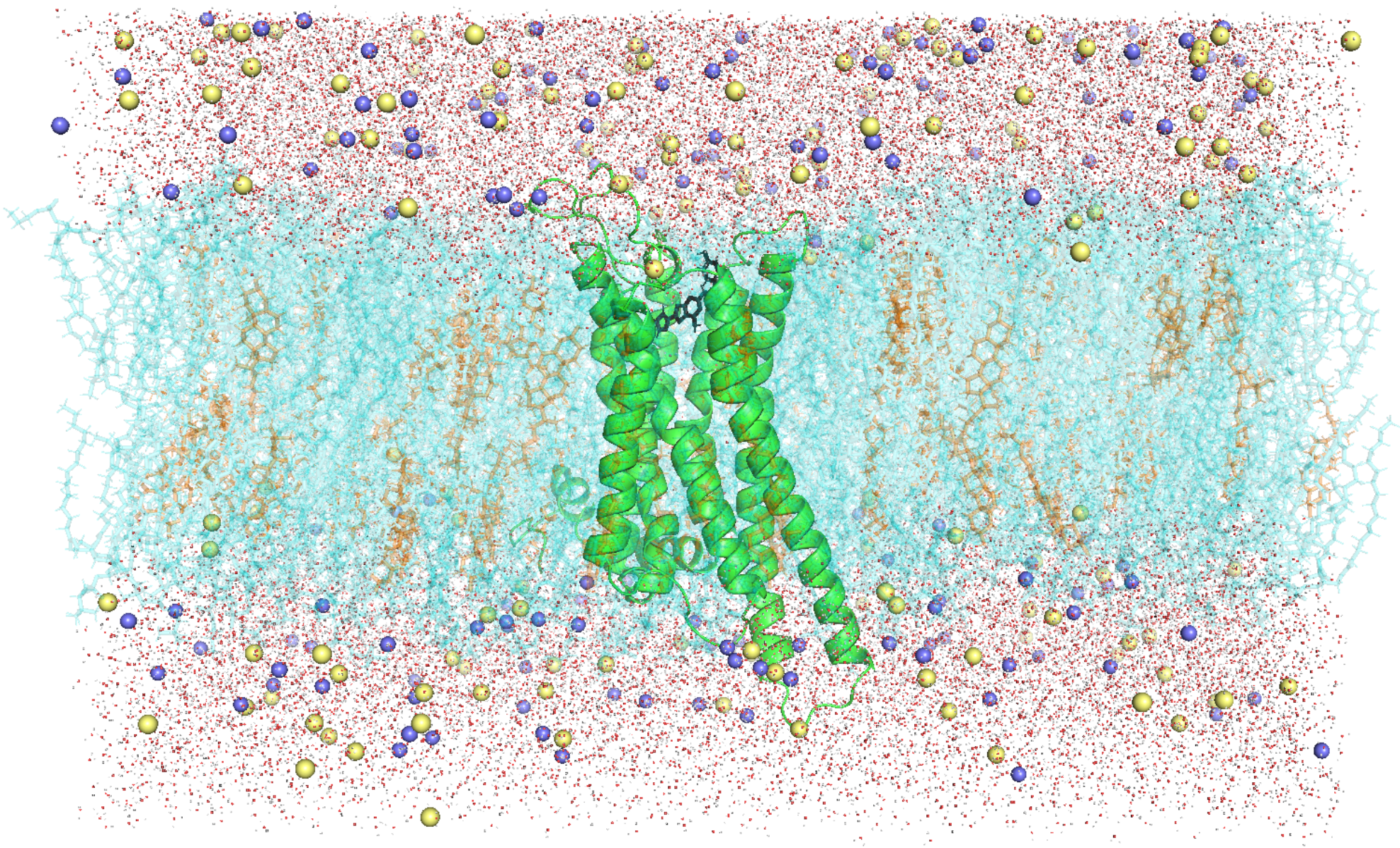

**Figure 3:** The system used in the benchmarks of the MoP implementation in GROMACS. The neuronal adenosine receptor type 2A (green ribbon) in complex with its ZM241385 inhibitor (blue), embedded in a lipid bilayer and fully solvated in explicit water.

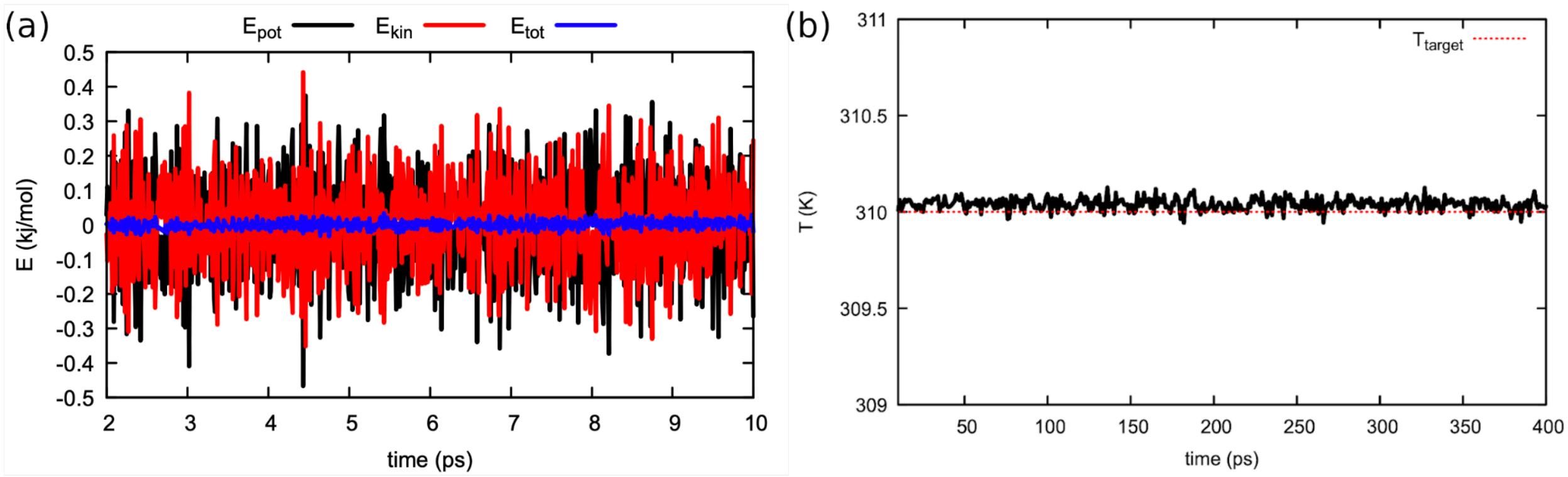

**Figure 4:** (a) NVE test: The black, red and blue lines report, respectively, the potential, kinetic and total energy (shifted relative to their average values). (b) NVT test: The black curve reports the instantaneous temperature as a function of time in a canonical simulation performed at the target temperature $T_{\text{target}} = 310$ K .

**3. Weak scaling tests**

The PMD algorithm represents an ideal weak scaling application, where one would like to simulate the longer possible polymer (increasing the number of beads) without affecting the time per PMD step by increasing the amount of computational resources (increasing number of nodes).

To demonstrate the efficiency of our parallel implementation in GROMACS, we have performed weak scaling tests using our model H2A receptor on the pre-exascale JUWELS Booster supercomputer at the Juelich Supercomputing Center [25]. JUWELS Booster consists of 936 compute nodes. Each node hosts AMD EPYC 7402 processors organized in 2 sockets with 24 cores per socket for a total of 24×2 = 48 physical cores per node. Each node is also equipped with 4 NVIDIA A100 GPUs.

Preliminary tests showed that the optimal configuration for our case study corresponds to offloading the force calculations of each bead to a separate GPU. We have therefore used this configuration for all our weak scaling tests. This corresponds to running simulations using 4 MPI tasks/node (one per GPU) and 6 OpenMP threads per task. In Table I we report the polymer sizes and the corresponding number of Booster nodes and GPUs used in our tests.

For each polymer size, we have performed an NVT simulation consisting of 30,000 MD steps. For a fair evaluation of the performance, we have ensured that all the internal, on-the-fly load-balance optimizations are completed before measuring the wall-time required to perform the final 5,000 steps.

Two sets of simulations have been performed. In the first one, we have performed unbiased PMD simulations. In the second one, we have performed metadynamics simulations using our PLUMED interface. As a CV, we have used the end-to-end distance defined in equation (10), where we chose s(**R**) to be the distance between the center of mass of the ligand and the center of mass of the binding pocket. The bias deposition rate has been fixed to 500 MD steps after which a new Gaussian is deposited. This is a typical rate used in classical metadynamics simulations.

Figure 5 and 6 shows the results of the weak scaling tests. For the case of unbiased PMD simulations, we observed an excellent weak scaling parallel efficiency above 70% up to the largest number of 3,500 GPUs used. This result is a direct consequence of our efficient MPMD parallel implementation of the algorithm that only makes use of point-to-point communication between adjacent beads. When performing metadynamics simulations, the weak scaling parallel performance is somewhat reduced, which is attributed to the additional communication overhead required to evaluate the CV.

| Polymer size | Nodes | GPUs |
|---|---|---|
| 4 | 1 | 4 |
| 400 | 100 | 400 |
| 800 | 200 | 800 |
| 1,200 | 300 | 1,200 |
| 1,600 | 400 | 1,600 |
| 2,000 | 500 | 2,000 |
| 2,400 | 600 | 2,400 |
| 2,800 | 700 | 2,800 |
| 3,200 | 800 | 3,200 |
| 3,500 | 875 | 3,500 |

**Table I.** The table reports the size of the polymer and the corresponding JUWELS Booster resources used for the weak scaling tests of Figures 5 and 6.

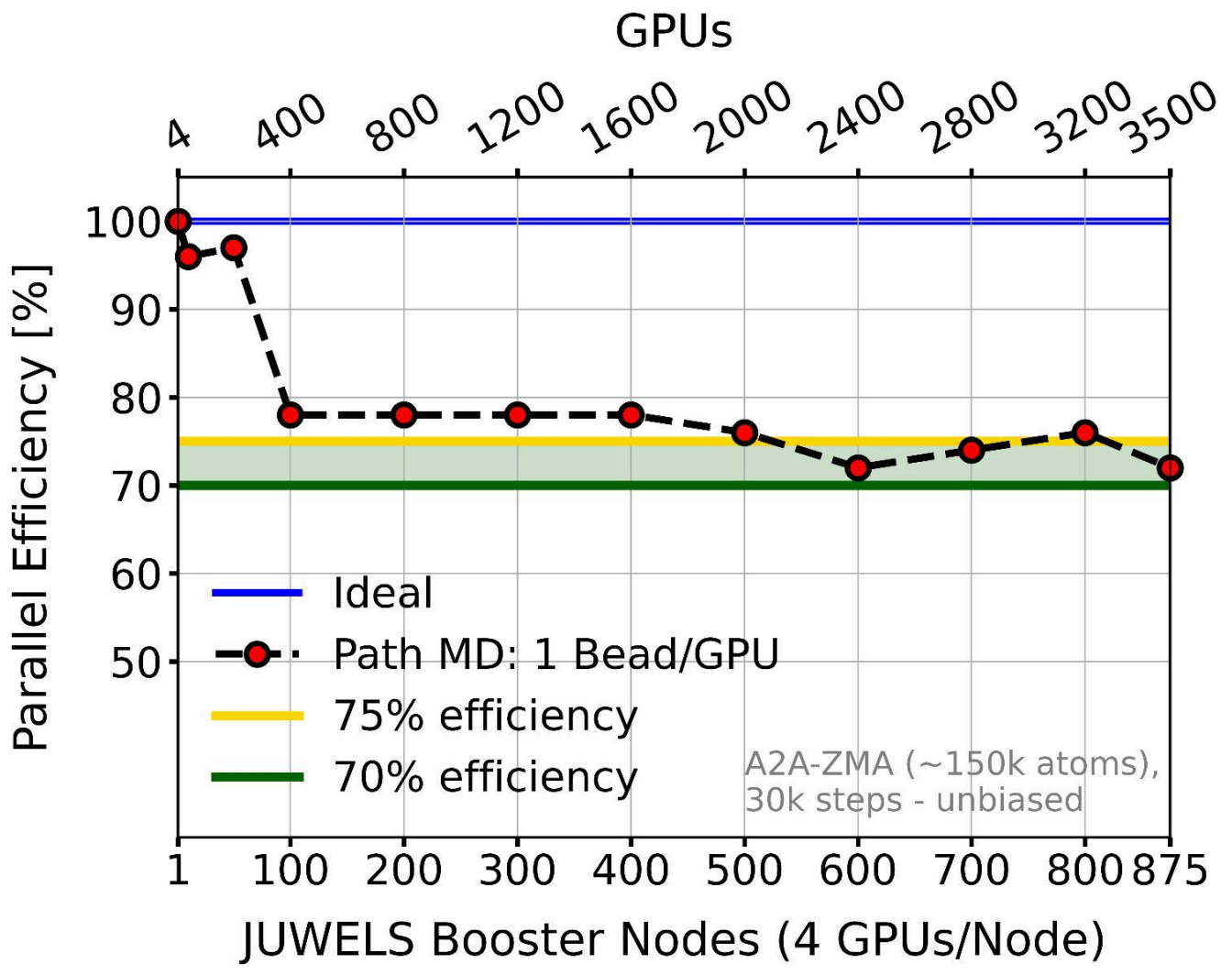

**Figure 5.** Weak scaling parallel efficiency of the path MD algorithm on the pre-exascale JUWELS Booster supercomputer.

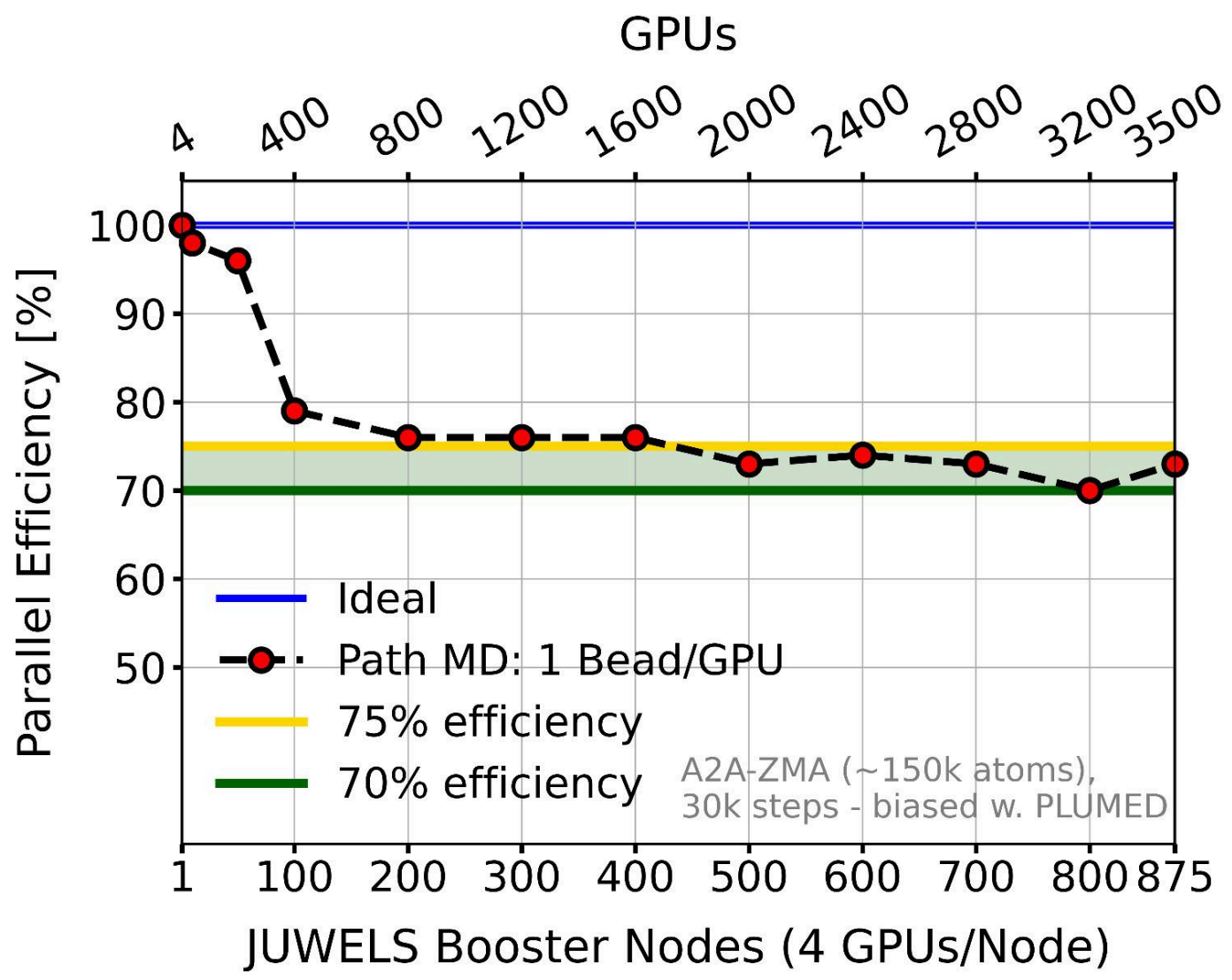

**Figure 6.** Weak scaling parallel efficiency of metadynamics of paths on the pre-exascale JUWELS Booster supercomputer.

## 4. Conclusion and outlook

We have presented a parallel implementation of the metadynamics of paths algorithm [1] in the GROMACS code for classical biomolecular simulations and we have demonstrated excellent weak scaling parallel efficiency above 70 % up to 3,500 GPUs on the pre-exascale JUWELS Booster machine [25] while considering as a test case a fairly large membrane protein. These promising results show that the approach is ready to take full advantage of modern accelerated computer architectures to investigate rare events in complex systems, such as drugs binding to their target neuroreceptors.

## Acknowledgements

This research was partly supported by the European Union's HORIZON MSCA Doctoral Networks programme, under Grant Agreement No. 101072344, project AQTIVATE (Advanced comput- ing, QuanTum algorIthms and data-driVen Approaches for Science, Technology, and Engineering). The project received funding from the Helmholtz European Partnering program ("Innovative high-performance computing approaches for molecular neuromedicine"). The authors gratefully acknowledge the Gauss Centre for Supercomputing e.V. (www.gauss-centre.eu) for funding this project by providing computing time through the John von Neumann Institute for Computing (NIC) on the GCS Supercomputer JUWELS [14] at Jülich Supercomputing Centre (JSC). DM and NM especially acknowledge M. Abraham for his guidance with the implementation in GROMACS and for all the helpful discussions.

## Bibliography

[1] D. Mandelli, B. Hirshberg, and M. Parrinello, "Metadynamics of Paths," *Phys. Rev. Lett.*,

vol. 125, no. 2, p. 026001, Jul. 2020, doi: 10.1103/PhysRevLett.125.026001.

[2] C. Dellago, P. G. Bolhuis, F. S. Csajka, and D. Chandler, “Transition path sampling and the calculation of rate constants,” *J. Chem. Phys.*, vol. 108, no. 5, pp. 1964–1977, Feb. 1998, doi: 10.1063/1.475562.

[3] A. Laio and M. Parrinello, “Escaping free-energy minima,” *Proc. Natl. Acad. Sci.*, vol. 99, no. 20, pp. 12562–12566, Oct. 2002, doi: 10.1073/pnas.202427399.

[4] G. Bussi, A. Laio, and M. Parrinello, “Equilibrium Free Energies from Nonequilibrium Metadynamics,” *Phys. Rev. Lett.*, vol. 96, no. 9, p. 090601, Mar. 2006, doi: 10.1103/PhysRevLett.96.090601.

[5] A. Barducci, G. Bussi, and M. Parrinello, “Well-Tempered Metadynamics: A Smoothly Converging and Tunable Free-Energy Method,” *Phys. Rev. Lett.*, vol. 100, no. 2, p. 020603, Jan. 2008, doi: 10.1103/PhysRevLett.100.020603.

[6] J. F. Dama, M. Parrinello, and G. A. Voth, “Well-Tempered Metadynamics Converges Asymptotically,” *Phys. Rev. Lett.*, vol. 112, no. 24, p. 240602, Jun. 2014, doi: 10.1103/PhysRevLett.112.240602.

[7] M. Parrinello and A. Rahman, “Study of an F center in molten KCl,” *J. Chem. Phys.*, vol. 80, no. 2, pp. 860–867, 1984, doi: 10.1063/1.446740.

[8] A. Putrino, D. Sebastiani, and M. Parrinello, “Generalized variational density functional perturbation theory,” *J. Chem. Phys.*, vol. 113, no. 17, pp. 7102–7109, Nov. 2000, doi: 10.1063/1.1312830.

[9] V. Kapil, J. Behler, and M. Ceriotti, “High order path integrals made easy,” *J. Chem. Phys.*, vol. 145, no. 23, p. 234103, Dec. 2016, doi: 10.1063/1.4971438.

[10] O. Valsson, P. Tiwary, and M. Parrinello, “Enhancing Important Fluctuations: Rare Events and Metadynamics from a Conceptual Viewpoint,” *Annu. Rev. Phys. Chem.*, vol. 67, no. 1, pp. 159–184, May 2016, doi: 10.1146/annurev-physchem-040215-112229.

[11] M. Bonomi, A. Barducci, and M. Parrinello, “Reconstructing the equilibrium Boltzmann distribution from well-tempered metadynamics,” *J. Comput. Chem.*, vol. 30, no. 11, pp. 1615–1621, Aug. 2009, doi: 10.1002/jcc.21305.

[12] D. Branduardi, G. Bussi, and M. Parrinello, “Metadynamics with Adaptive Gaussians,” *J. Chem. Theory Comput.*, vol. 8, no. 7, pp. 2247–2254, Jul. 2012, doi: 10.1021/ct3002464.

[13] P. Tiwary and M. Parrinello, “A Time-Independent Free Energy Estimator for Metadynamics,” *J. Phys. Chem. B*, vol. 119, no. 3, pp. 736–742, Jan. 2015, doi: 10.1021/jp504920s.

[14] L. Mones, N. Bernstein, and G. Csányi, “Exploration, Sampling, And Reconstruction of Free Energy Surfaces with Gaussian Process Regression,” *J. Chem. Theory Comput.*, vol. 12, no. 10, pp. 5100–5110, Oct. 2016, doi: 10.1021/acs.jctc.6b00553.

[15] V. Marinova and M. Salvalaglio, “Time-independent free energies from metadynamics via mean force integration,” *J. Chem. Phys.*, vol. 151, no. 16, p. 164115, Oct. 2019, doi: 10.1063/1.5123498.

[16] M. Invernizzi and M. Parrinello, “Rethinking Metadynamics: From Bias Potentials to Probability Distributions,” *J. Phys. Chem. Lett.*, vol. 11, no. 7, pp. 2731–2736, Apr. 2020, doi: 10.1021/acs.jpclett.0c00497.

[17] F. Giberti, B. Cheng, G. A. Tribello, and M. Ceriotti, “Iterative Unbiasing of Quasi-Equilibrium Sampling,” *J. Chem. Theory Comput.*, vol. 16, no. 1, pp. 100–107, Jan. 2020, doi: 10.1021/acs.jctc.9b00907.

[18] L. Müllender, A. Rizzi, M. Parrinello, P. Carloni, and D. Mandelli, “Effective data-driven collective variables for free energy calculations from metadynamics of paths,” *PNAS Nexus*, vol. 3, no. 4, pp. 1–9, Mar. 2024, doi: 10.1093/pnasnexus/pgae159.

[19] M. J. Abraham *et al.*, “GROMACS: High performance molecular simulations through multi-level parallelism from laptops to supercomputers,” *SoftwareX*, vol. 1–2, pp. 19–25, Sep. 2015, doi: 10.1016/j.softx.2015.06.001.

[20] C. Kutzner, S. Páll, M. Fechner, A. Esztermann, B. L. Groot, and H. Grubmüller, "More bang for your buck: Improved use of GPU nodes for GROMACS 2018," *J. Comput. Chem.*, vol. 40, no. 27, pp. 2418–2431, Oct. 2019, doi: 10.1002/jcc.26011.

[21] S. Páll *et al.*, "Heterogeneous parallelization and acceleration of molecular dynamics simulations in GROMACS," *J. Chem. Phys.*, vol. 153, no. 13, p. 134110, Oct. 2020, doi: 10.1063/5.0018516.

[22] M. Bonomi *et al.*, "PLUMED: A portable plugin for free-energy calculations with molecular dynamics," *Comput. Phys. Commun.*, vol. 180, no. 10, pp. 1961–1972, Oct. 2009, doi: 10.1016/j.cpc.2009.05.011.

[23] G. A. Tribello, M. Bonomi, D. Branduardi, C. Camilloni, and G. Bussi, "PLUMED 2: New feathers for an old bird," *Comput. Phys. Commun.*, vol. 185, no. 2, pp. 604–613, Feb. 2014, doi: 10.1016/j.cpc.2013.09.018.

[24] N. Malapally, D. Mandelli, and P. Carloni, "(In preparation)," in *Computational Modeling of Biomolecular Interactions: Methods and Applications*, Y. Miao, Ed., Wiley, 2025.

[25] D. Alvarez, "JUWELS Cluster and Booster: Exascale Pathfinder with Modular Supercomputing Architecture at Juelich Supercomputing Centre," *J. Large-Scale Res. Facil. JLSRF*, vol. 7, p. A183, Oct. 2021, doi: 10.17815/jlsrf-7-183.

[26] R. Cao, A. Giorgetti, A. Bauer, B. Neumaier, G. Rossetti, and P. Carloni, "Role of Extracellular Loops and Membrane Lipids for Ligand Recognition in the Neuronal Adenosine Receptor Type 2A: An Enhanced Sampling Simulation Study," *Molecules*, vol. 23, no. 10, p. 2616, Oct. 2018, doi: 10.3390/molecules23102616.

# Appendix A. Methods

### A.1 Standard MD simulations

Human adenosine receptor type 2A (hA2AR) is a class A GPCR, composed of 7 transmembrane helices (H1-H7) and a helix lying at the membrane-cytoplasm interface (H8). hA2AR has a high pharmacological relevance: highly localized in the striatum of the brain, it is considered a promising drug target for combating Parkinson’s disease.
Preliminary, standard molecular dynamics (MD) simulations on the $hA_{2A}R$ in complex with its high affinity antagonist ZMA were carried out using GROMACS v2023 on JUWELS Cluster module of the Jülich Supercomputing Center (JSC). The initial MD configuration was taken from a snapshot of our previous MD simulation of ZMA/$hA_{2A}R$ complex [26], embedded the receptor in a membrane of 42% POPC, 34% POPE and 25% of cholesterol molecules, mimicking the ratio among the three components in human cellular plasma membranes. The selected snapshot shows ZMA bound to the receptor’s orthosteric binding site. The system was inserted in a simulation box of size (14.3 x 10.8 x 9.6) nm, including 248 POPC lipids, 204 POPE lipids, and 141 cholesterol molecules, and solvated with water and 150 mM NaCl. The final system consists of 151,850 atoms. The AMBER99SB-ILDN force fields, the Slipids, the TIP3P force fields were used for the protein and ions, the lipids, and the water molecules respectively. The General Amber force field (GAFF) parameters were used for ZMA, along with the RESP atomic charge using Gaussian 09 with the HF-6-31G* basis set. The same computational protocol as in Ref. [26] was followed to equilibrate the system. Specifically, keeping the position restraints for the protein and the ligand, we have performed 20 ns simulation in the NVT ensemble at 310 K; after this step, we have released the ligand, and the system underwent a total of 60 ns simulation in the NPT ensemble. Initially, the system was equilibrated using the Berendsen barostat with a 1 ps coupling constant and semi-isotropic pressure coupling type to maintain pressure at 1 bar. Isothermal compressibility was set to $4.5 \times 10^{-5}$ $bar^{-1}$. Temperature was set to 310 K and maintained using the Nose-Hoover thermostat with a 0.5 ps coupling constant. Explicit temperature groups were defined, with protein and ligand in one group, lipid atoms in one group, and solvent and ions in another, to reduce temperature-induced artifacts. After 5 ns of simulation, constant temperature and pressure conditions were achieved via independently coupling protein, lipids, solvent and ions to Nosè-Hoover thermostat and Andersen-Parrinello-Rahman barostat, to ensure a reliable maintenance of the isothermal–isobaric ensemble. A coupling constant of 0.5 ps was used for maintaining temperature at 310 K and 1 ps for maintaining pressure at 1 bar with semi-isotropic coupling. The Particle Mesh Ewald method was used to treat the long-range electrostatic interaction with a real space cutoff of 1.2 nm. A 1.2 nm cutoff was used for the short-range non-bonded interaction. A time- step of 2 fs was set. The LINCS algorithm was applied to constrain all bonds involving hydrogen atoms, and trajectory frames were saved every 2 ps.

### A.2 PMD and MoP simulations

All the PMD and MoP simulations have been performed using the same force field discussed in Section A.1. For the NVE and NVT validation runs, we have used a polymer made up of N=512 beads. The assumption behind the MoP theory is that the polymer represents a Brownian trajectory of duration, where the *n*-th bead corresponds to the configuration at time $n\Delta t$ along the trajectory. Hence, starting from the equilibrated structure obtained at the end of the standard MD simulations, we have performed standard Brownian Dynamics simulations (as implemented in GROMACS) for $10^6$ steps and saved the last 512 snapshots (saving configurations every time step). These were used as the starting configurations for the polymer beads in PMD runs.
In PMD NVE tests, a velocity Verlet integrator was used with an integration time step of $\Delta t = 1\,fs$. In PMD NVT, the same time step was used in combination with the Langevin thermostat, as implemented in GROMACS, using a damping coefficient of $\nu = 0.25\,fs^{-1}$.
We followed a similar procedure to set up the initial configuration for the weak scaling tests with different polymer sizes (see Table I). The latter consisted of 25,000 PMD or MoP steps in the NVT ensemble, using the same setup discussed above.